\documentclass[12pt,prd,tightenlines,nofootinbib,showpacs,showkeys]{revtex4}
\newcommand{\be}{\begin{equation}}
\newcommand{\ee}{\end{equation}}
\usepackage{bm}
\usepackage{graphics}
\usepackage{rotating}
\usepackage{epsfig}
\begin{document}
\title{\begin{flushright}{\rm\normalsize SSU-HEP-11/05\\[5mm]}\end{flushright}
Relativistic description of the double P-wave \\
charmonium production in $e^+e^-$ annihilation}
\author{\firstname{A.P.} \surname{Martynenko}}
\affiliation{Samara State University, Pavlov Street 1, 443011, Samara, Russia}
\affiliation{Samara State Aerospace University named after S.P. Korolyov, Moskovskoye Shosse 34, 443086,
Samara, Russia}
\author{\firstname{A.M.} \surname{Trunin}}
\affiliation{Samara State Aerospace University named after S.P. Korolyov, Moskovskoye Shosse 34, 443086,
Samara, Russia}

\begin{abstract}
On the basis of perturbative QCD and the relativistic quark model we
calculate relativistic and bound state corrections in the production
processes of a pair of ${\cal P}$-wave charmonium states.
Relativistic factors in the production amplitude connected with the
relative motion of heavy quarks and the transformation law of the
bound state wave function to the reference frame of the moving
${\cal P}$-wave mesons are taken into account. For the gluon and quark
propagators entering the production vertex function we use a
truncated expansion in the ratio of the relative quark momenta to
the center-of-mass energy $\sqrt{s}$ up to the second order.
Relativistic corrections to the quark bound state wave functions in
the rest frame are considered by means of the Breit-like potential.
It turns out that the examined effects change essentially the
nonrelativistic results of the cross section for the reaction
$e^++e^-\to h_c+\chi_{cJ}$ at the center-of-mass
energy $\sqrt{s}=10.6$ GeV.
\end{abstract}



\maketitle

\section{Introduction}

The large value of the exclusive double charmonium production cross section
measured at the Belle and BABAR experiments \cite{Belle,BaBar}
reveals definite problems in the theoretical description of these
processes \cite{BL1,Chao,Qiao}. Many theoretical efforts were made
in order to improve the calculation of the production cross section
$e^++e^-\to J/\Psi+\eta_c$. They included the analysis of other
production mechanisms for the state $J/\Psi+\eta_c$ \cite{BLB,BGL}
and the calculation of different corrections which could change
essentially the initial nonrelativistic result
\cite{Chao1,Ma,BC,BLL,ZGC,Bodwin2,EM2006,Ji,He,AVB,Bodwin4}. Despite
the evident successes achieved on the basis of nonrelativistic quantum
chromodynamics (NRQCD), the light cone
method, quark potential models for correcting the discrepancy
between the theory and experiment, the double charmonium production
in $e^+e^-$ annihilation remains an interesting task. On the one
hand, there are other production processes of the
${\cal P}$- and ${\cal D}$-wave charmonium states
which can be investigated in the same way as the production of ${\cal S}$-wave
states. Recently the Belle and BABAR collaborations
discovered new charmonium-like states in $e^+e^-$ annihilation \cite{pahlova,brambilla-2011}.
The nature of these numerous resonances remains unclear to the present. Some
of them are considered as a ${\cal P}$- and ${\cal D}$-wave excitations in the system ($c\bar c$).
On the other hand, the variety of the
used approaches and the model parameters in this problem raises the
question about the comparison of the obtained results
that will lead to a better understanding of the quark-gluon dynamics and
different mechanisms of the charmonium production.
Two sources of the
changing of the nonrelativistic cross section for the double
charmonium production are revealed to the present: the radiative
corrections of order $O(\alpha_s)$ and relative motion of $c$-quarks
forming the bound states. An actual physical processes of the charmonium
production require formation of hadronic particles in final states
(bound states of a charm quark $c$ and a charm anti-quark $\bar c$), for
which quantum chromodynamics can not provide high precision description.
Further investigation of charmonia production can improve our understanding
of heavy quark production and the formation of quark bound states.

This work continues our study of the exclusive double
charmonium production in $e^+e^-$ annihilation in the case of a pure
${\cal P}$-wave $(c\bar c)$ quarkonium on the basis of a
relativistic quark model (RQM) \cite{EM2006,EFGM2009,EM2010,apm2005,rqm5}.
Note that the term RQM specifies the approach in which the systematic
account of corrections connected with the relative motion of heavy quarks
can be performed. The relativistic quark
model provides the solution in many tasks of heavy quark physics.
It uses a number of perturbative and nonperturbative parameters
entering in the quark interaction operator. All observables can be
expressed in terms of these parameters. In this way we can check the
predictions of any quark model and draw a conclusion about its
successfulness. At the same time the existence of a large number of different
quark models which are sometimes very complicated for the practical use
put a question about the elaboration of the unified model containing
generally accepted structural elements.
Another approach to the heavy quark physics which does not contain the ambiguities
of the quark models was formulated in \cite{BBL}. As any other model
of strong interactions of quarks and gluons the approach of NRQCD introduces
in the theory a large number of matrix elements parameterizing nonperturbative
dynamics of quarks. To a certain extent the microscopic picture of the
quark-gluon interaction resident in quark models is changed by the global
picture operating with the numerous nonperturbative matrix elements. The
improved determination of color-singlet NRQCD matrix elements for $S$-wave
charmonium is presented in \cite{bodwin_anl}. Their study evidently shows
that the account of relative order $v^2$ corrections significantly increases
the values of the matrix elements of leading order in $v$.
The correspondence between parameters of quark models and NRQCD which can
be established, opens the way for better understanding of quark-gluon interactions
at small distances. In this sense both approaches complement each other and
could reveal new aspects of color dynamics of quarks and gluons. Thus, the aim of
this study consists in the extension of relativistic approach to the
quarkonium production from Refs.\cite{EM2006,EFGM2009,EM2010} on the
processes $e^++e^-\to h_c+\chi_{cJ}$ and determination of the interrelationship with
the predictions of NRQCD.

\section{General formalism}

We investigate the quarkonium production in the lowest-order perturbative quantum chromodynamics.
The usual color-singlet mechanism is considered as a basic one for the pair charmonium production.
We analyze the reactions $e^++e^-\to h_c+\chi_{cJ}$, where the final state consists of
a pair of ${\cal P}$-wave ($\chi_{c0}$, $\chi_{c1}$, $\chi_{c2}$) and $h_c$ charm mesons.
The diagrams that give contributions to the amplitude of these processes in
leading order of the QCD coupling constant $\alpha_s$ are presented
in Fig.1. Two other diagrams can be obtained by corresponding
permutations. There are two stages of the production process. In the
first stage, which is described by perturbative QCD, the virtual
photon $\gamma^\ast$ produces four heavy $c$-quarks and $\bar
c$-antiquarks with the following four-momenta:

\begin{figure}
\centering
\includegraphics[width=5.cm]{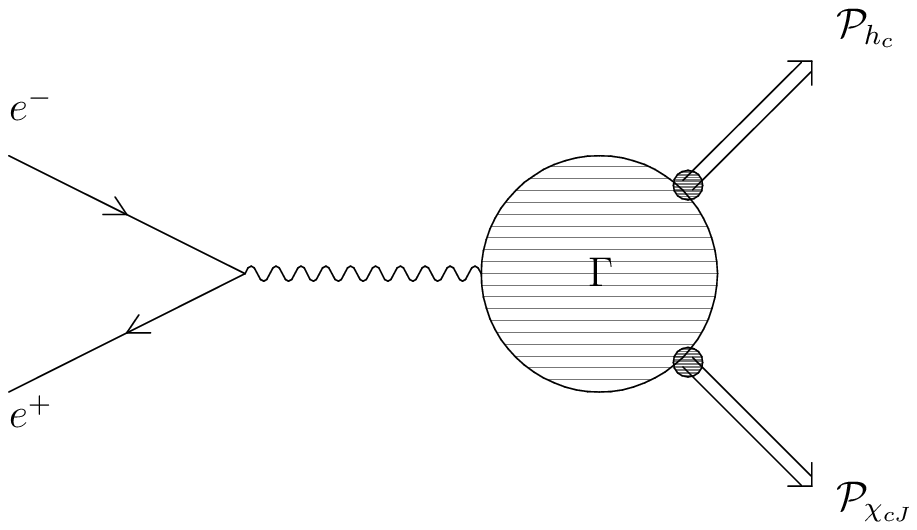}\hspace*{0.4cm}
\includegraphics[width=11cm]{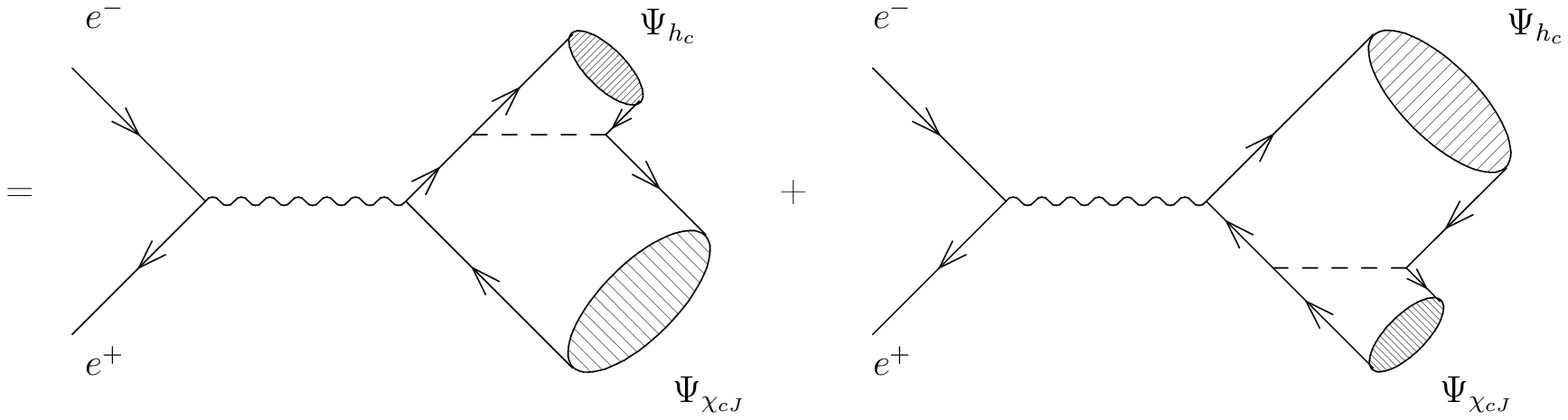}
\caption{The production amplitude of a pair of ${\cal P}$-wave
charmonium states in $e^+e^-$ annihilation. ${\cal P}_{h_c}$ denotes the
${\cal P}$-wave meson $h_c$ and ${\cal P}_{\chi_{cJ}}$ denotes the ${\cal P}$-wave meson $\chi_{cJ}$. The wavy line shows
the virtual photon and the dashed line corresponds to the gluon.
$\Gamma$ is the production vertex function.}
\end{figure}

\begin{equation}
p_{1,2}=\frac{1}{2}P\pm p,~~(p\cdot P)=0;~~q_{1,2}=\frac{1}{2}Q\pm q,~~(q\cdot Q)=0,
\end{equation}
where $P(Q)$ are the total four-momenta, $p=L_P(0,{\bf p})$,
$q=L_P(0,{\bf q})$ are the relative four-momenta obtained from the
rest frame four-momenta $(0,{\bf p})$ and $(0,{\bf q})$ by the
Lorentz transformation to the system moving with the momenta $P$, $Q$.
The momenta $p_{1,2}$ of the heavy quark $c$ and antiquark $\bar c$ are not on
the mass shell: $p_{1,2}^2=P^2/4-{\bf p}^2=M^2/4-{\bf p}^2\not= m^2$. Relation (1)
describes the symmetrical escape of the $c$-quark and $\bar c$-antiquark from
the mass shell. In the second nonperturbative stage, quark-antiquark pairs form the
final mesons.

Let consider the production amplitude of the ${\cal P}$-wave vector state
$h_c$ and ${\cal P}$-wave states $\chi_{cJ}$ ($J=0,1,2$), which can be
presented in the form \cite{rqm5,EM2006,EM2010}:
\begin{equation}
{\cal M}(p_-,p_+,P,Q)=\frac{8\pi^2\alpha\alpha_{s}(4m^2){\cal
Q}_c}{3s}\bar v(p_+)\gamma^\beta u(p_-)\int\frac{d{\bf
p}}{(2\pi)^3}\int\frac{d{\bf q}}{(2\pi)^3}\times
\end{equation}
\begin{displaymath}
\times Sp\left\{\Psi^{\cal
P}_{h_c}(p,P)\Gamma_1^{\beta\nu}(p,q,P,Q)\Psi^{\cal
P}_{\chi_{cJ}}(q,Q)\gamma_\nu+\Psi^{\cal
P}_{\chi_{cJ}}(q,Q)\Gamma_2^{\beta\nu}(p,q,P,Q)\Psi^{\cal P}_{h_c}(p,P)\gamma_\nu
\right\},
\end{displaymath}
where a superscript ${\cal P}$ indicates the ${\cal P}$-wave meson, $\alpha_{s}(4m^2)$
is the QCD coupling constant, $\alpha$ is the
fine structure constant and ${\cal Q}_c$ is the $c$-quark electric
charge, $\Gamma_{1,2}$ are the vertex functions defined below.
The production processes $e^++e^-\to h_c+\chi_{cJ}$ contain the quark
bound states. The transition of free quarks to the $(c\bar c)$ mesons
is described by specific wave functions. The relativistic
${\cal P}$-wave functions of the bound quarks
$\Psi^{\cal P}$ accounting for the transformation from the
rest frame to the moving one with four momenta $P,Q$, are
\begin{eqnarray}
\Psi^{\cal P}_{h_c}(p,P)&=&\frac{\Psi_0^{h_c}({\bf
p})}{\left[\frac{\epsilon(p)}{m}
\frac{(\epsilon(p)+m)}{2m}\right]}\left[\frac{\hat v_1-1}{2}+\hat
v_1\frac{{\bf p}^2}{2m(\epsilon(p)+ m)}-\frac{\hat{p}}{2m}\right]\cr
&&\times\gamma_5(1+\hat v_1) \left[\frac{\hat
v_1+1}{2}+\hat v_1\frac{{\bf p}^2}{2m(\epsilon(p)+
m)}+\frac{\hat{p}}{2m}\right],
\end{eqnarray}
\begin{eqnarray}
\Psi^{\cal P}_{\chi_{cJ}}(q,Q)&=&\frac{\Psi_0^{\chi_{cJ}}({\bf q})}
{\left[\frac{\epsilon(q)}{m}\frac{(\epsilon(q)+m)}{2m}\right]}
\left[\frac{\hat v_2-1}{2}+\hat v_2\frac{{\bf q}^2}{2m(\epsilon(q)+
m)}+\frac{\hat{q}}{2m}\right]\cr &&\times\hat{\varepsilon}_{\cal
P}^\ast(Q,S_z)(1+\hat v_2) \left[\frac{\hat v_2+1}{2}+\hat
v_2\frac{{\bf q}^2}{2m(\epsilon(q)+ m)}-\frac{\hat{q}}{2m}\right],
\end{eqnarray}
where the hat is a notation for the contraction of the four vector with
the Dirac matrices, $v_1=P/M_{h_c}$, $v_2=Q/M_{\chi_{cJ}}$;
$\varepsilon_{\cal P}(Q,S_z)$ is the polarization vector of the
spin-triplet state $\chi_{cJ}$, $\epsilon(p)=\sqrt{p^2+m^2}$ and $m$
is the $c$-quark mass. The relativistic functions (3)-(4) and the vertex functions $\Gamma_{1,2}$
do not contain the $\delta ({\bf p}^2-M^2/4+m^2)$.
More complicated factor including the bound state wave function in the rest frame presented
in Eqs.(3) and (4) plays the role of the $\delta$-function.
This means that instead of the substitutions $M_{h_c}=2\epsilon({\bf p})$ and
$M_{\chi_{cJ}}=2\epsilon({\bf q})$ in the production amplitude we carry out the
integration over the quark relative momenta ${\bf p}$ and ${\bf q}$.
The amplitude (2) is projected onto a color
singlet state by replacing $v_i(0)\bar u_k(0)$ with a projection
operator of the form $v_i(0)\bar u_k(0)=\delta_{ik}/\sqrt{3}$.
The relativistic wave functions in Eqs.(3),
(4) are equal to the product of the wave functions in the rest frame
$\Psi_0^{\cal P}$ and the spin projection operators that are
accurate at all orders in $|{\bf p}|/m$ \cite{rqm5,EM2006}. The
expression of the spin projector in a slightly different form has
been derived primarily in \cite{Bodwin2002} in the framework of NRQCD.
Our derivation of relations (3), (4) accounts for the transformation
law of the bound state wave functions from the rest frame to the
moving one with four momenta $P$ and $Q$. This transformation law
was discussed in the Bethe-Salpeter approach in \cite{BP} and in the
quasipotential method in \cite{F1973}. We use the last one and write
the necessary transformation as follows:
\begin{equation}
\Psi_{P}^{\rho\omega}({\bf p})=D_1^{1/2,~\rho\alpha}(R^W_{L_{P}})
D_2^{1/2,~\omega\beta}(R^W_{L_{P}})\Psi_{0}^{\alpha\beta}({\bf p}),
\end{equation}
\begin{displaymath}
\bar\Psi_{P}^{\lambda\sigma}({\bf p})
=\bar\Psi^{\varepsilon\tau}_{0}({\bf p})D_1^{+~1/2,~\varepsilon
\lambda}(R^W_{L_{P}})D_2^{+~1/2,~\tau\sigma}(R^W_{L_{P}}),
\end{displaymath}
where $R^W$ is the Wigner rotation, $L_{P}$ is the Lorentz boost
from the meson rest frame to a moving one, and the rotation matrix
$D^{1/2}(R)$ is defined by
\begin{equation}
{1 \ \ \,0\choose 0 \ \ \,1}D^{1/2}_{1,2}(R^W_{L_{P}})= S^{-1}({\bf
p}_{1,2})S({\bf P})S({\bf p}),
\end{equation}
where the explicit form for the Lorentz transformation matrix of the
four-spinor is
\begin{equation}
S({\bf p})=\sqrt{\frac{\epsilon(p)+m}{2m}}\left(1+\frac{(\bm{\alpha}
{\bf p})} {\epsilon(p)+m}\right).
\end{equation}

We omit here the intermediate expressions giving rise to our final relations (2)-(4)
\cite{EM2006,EFGM2009}. The presence of the $\delta (p\cdot P)$ function
allows to make the integration over relative energy $p^0$ if we write the initial
production amplitude as a convolution of the truncated amplitude with two
Bethe-Salpeter (BS) meson wave functions. In the rest frame of the bound state the condition
$p^0=0$ allows to eliminate the relative energy
from the BS wave function. The BS wave function satisfies a two-body bound state equation
which is very complicated and has no known solution. A way to deal with this problem
is to find a soluble lowest-order equation containing the main physical properties
of the exact equation and develop a perturbation theory. For this purpose we continue
to work in three-dimensional quasipotential approach. In this framework the double
charmonium production amplitude (2) can be written initially as a product of the production
vertex function $\Gamma_{1,2}$ projected onto the positive energy states by means of the Dirac
bispinors (free quark wave functions) and a bound state quasipotential wave functions
describing the ${\cal P}$-wave mesons in the reference frames moving with four momenta $P,Q$.
Further transformations use the known transformation law for the bound state wave
functions to the rest frame (5). The physical
interpretation of the double charmonium production amplitude is the following:
we have a complicated transition of two heavy $c$-quark and $\bar c$-antiquark
which were produced in $e^+e^-$-annihilation outside the mass shell and their
subsequent evolution firstly on the mass shell (free Dirac bispinors) and then to the
quark bound states. In the spin projectors we have
${\bf p}^2\not=M^2/4-m^2$ just the same as in the vertex production functions $\Gamma_{1,2}$.
We can not say exactly whether the charm quarks are on-shell or not in the spin
projectors (3)-(4) because we should consider these structures as a transition form factors for
the heavy quarks from the free states to the bound states.

At leading order in $\alpha_s$ the vertex functions
$\Gamma_{1,2}^{\beta\nu}(p,P;q,Q)$ can be written as
\begin{equation}
\Gamma_1^{\beta\nu}(p,P;q,Q)= \gamma_\mu\frac{(\hat l-\hat
q_1+m)}{(l-q_1)^2-m^2+i\epsilon} \gamma_\beta D^{\mu\nu}(k_2)+
\gamma_\beta\frac{(\hat p_1-\hat l+m)}{(l-p_1)^2-m^2+i\epsilon}
\gamma_\mu D^{\mu\nu}(k_2),
\end{equation}
\begin{equation}
\Gamma_2^{\beta\nu}(p,P;q,Q)=\gamma_\beta\frac{(\hat q_2-\hat
l+m)}{(l-q_2)^2-m^2+i\epsilon} \gamma_\mu D^{\mu\nu}(k_1)+
\gamma_\mu\frac{(\hat l-\hat p_2+m)}{(l-p_2)^2-m^2+i\epsilon}
\gamma_\beta D^{\mu\nu}(k_1),
\end{equation}
where the gluon momenta are $k_1=p_1+q_1$, $k_2=p_2+q_2$ and
$l^2=s=(P+Q)^2=(p_-+p_+)^2$, $p_-$, $p_+$ are four momenta of the
electron and positron. The dependence on the relative momenta of
$c$-quarks is presented both in the gluon propagator $D_{\mu\nu}(k)$
and quark propagator as well as in the relativistic wave functions
(3), (4). Taking into account that the ratio of the relative quark
momenta $p$ and $q$ to the energy $\sqrt{s}$ is small, we expand the
inverse denominators of quark and gluon propagators as follows:
\begin{equation}
\frac{1}{(l-q_{1,2})^2-m^2}=\frac{2}{s}\left[1-\frac{2M_{h_c}^2-M_{\chi_{cJ}}^2-4m^2}{2s}-\frac{2q^2}{s}\pm\frac{4(lq)}
{s}+\frac{16(lq)^2}{s^2}+\cdots\right],
\end{equation}
\begin{equation}
\frac{1}{(l-p_{1,2})^2-m^2}=\frac{2}{s}\left[1-\frac{2M_{\chi_{cJ}}^2-M_{h_c}^2-4m^2}{2s}-\frac{2p^2}{s}\pm\frac{4(lp)}
{s}+\frac{16(lp)^2}{s^2}+\cdots\right],
\end{equation}
\begin{equation}
\frac{1}{k_{2,1}^2}=\frac{4}{s}\left[1-\frac{4(p^2+q^2+2pq)}{s}\pm\frac{4(lp+lq)}{s}+
\frac{16}{s^2}\left[(lp)^2+(lq)^2+2(lp)(lq)\right]+\cdots\right].
\end{equation}
In the expansions (10)-(12) we keep terms of third order in
relative momenta $p$ and $q$.
Substituting (10)-(12), (3)-(4) in (2) we preserve relativistic
factors entering the denominators of the relativistic wave functions
(3)-(4), but in the numerator of the amplitude (2) we take into
account corrections of third order in $|{\bf p}|/m$ and $|{\bf q}|/m$.
This provides the convergence of the
resulting momentum integrals. Then the angular integrals are
calculated using the following relations:
\begin{equation}
\int q_\mu\frac{\Psi_0^{\cal P}({\bf
q})}{\left[\frac{\epsilon(q)}{m}
\frac{(\epsilon(q)+m)}{2m}\right]}\frac{d{\bf
q}}{(2\pi)^3}=-i\varepsilon_{{\cal
P}\mu}(Q,L_z)\frac{1}{\pi\sqrt{6}}\int_0^\infty q^3 \frac{R_{\cal
P}(q)}{\left[\frac{\epsilon(q)}{m}
\frac{(\epsilon(q)+m)}{2m}\right]}dq,
\end{equation}
\begin{equation}
\int \frac{q_\alpha q_\beta q_\gamma\Psi_0^{\cal P}({\bf
q})}{\left[\frac{\epsilon(q)}{m}
\frac{(\epsilon(q)+m)}{2m}\right]}\frac{d{\bf
q}}{(2\pi)^3}=\frac{i}{5\pi\sqrt{6}}[\varepsilon_\gamma(Q,L_z)P_{\alpha\beta}+
\varepsilon_\alpha(Q,L_z)P_{\gamma\beta}+\varepsilon_\beta(Q,L_z)
P_{\alpha\gamma}]\int_0^\infty \frac{q^5R_{\cal
P}(q)}{\left[\frac{\epsilon(q)}{m}
\frac{(\epsilon(q)+m)}{2m}\right]}dq,
\end{equation}
where $P_{\alpha\beta}=(g_{\alpha\beta}-v_{2\alpha}v_{2\beta})$,
$R_{\cal P}(q)$ is the radial momentum wave
function of ${\cal P}$-wave charmonium states,
$\varepsilon_\mu(Q,L_z)$ is the polarization vector in orbital
space. The integrals in (13) and (14) are convergent due to the presence
of relativistic factors. In this work we do not make
expansions of all relativistic factors containing the relative momenta ${\bf p}$, ${\bf q}$
as in our paper \cite{EM2006}. This gives us an opportunity to calculate the corrections
of the second order in $|{\bf p}|/m$, $|{\bf q}|/m$ working with the convergent integrals.
Both approaches of the expansion can be used. In the first one we expand all factors,
immediately obtain the divergent integrals and should find additional arguments
to fix its numerical value \cite{Bodwin2}. On the second way, which we take in this work,
we have no the divergent integrals in the corrections of second order.
Undecomposed relativistic factors in (13)-(14) can be considered as a natural
cutoff of the momentum integrals. For a specific ${\cal P}$-wave
state, summing over $S_z$ and $L_z$ in the amplitude (2) can be
further simplified as \cite{Kuhn}
\begin{equation}
\sum_{S_z,L_z}\langle 1,L_z;1,S_z|J,J_z\rangle \varepsilon^\ast_{{\cal
P}\alpha}(Q,L_z)\varepsilon^\ast_{{\cal P}\beta}(Q,S_z)=
\cases{\frac{1}{\sqrt{3}}(g_{\alpha\beta}-v_{2\alpha}v_{2\beta}),&$J=0$,\cr
\frac{i}{\sqrt{2}}\epsilon_{\alpha\beta\sigma\rho}v_2^\sigma\varepsilon^{\ast\rho}(Q,J_z),&$J=1$,\cr
\varepsilon^\ast_{\alpha\beta}(Q,J_z),&$J=2$,\cr}
\end{equation}
where $\langle 1,L_z;1,S_z|J,J_z\rangle $ are the Clebsch-Gordon coefficients.
Calculating the trace in the amplitude (2) by means of expressions
(3)-(4), (8)-(9) and the system FORM \cite{FORM}, we find that the
tensor parts of four amplitudes describing the production of
${\cal P}$-wave charmonium states have the following structure:
\begin{equation}
S_{0,\beta}(h_c+\chi_{c0})=A_0\varepsilon_{\mu\nu\alpha\beta}v_1^{\mu}v_2^\nu
\varepsilon^{\ast\alpha}_{h_c},
\end{equation}
\begin{equation}
S_{1,\beta}(h_c+\chi_{c1})=B_1v_{1\beta}(v_1\cdot\varepsilon^\ast_{\chi_{c1}})(v_2\cdot
\varepsilon^\ast_{h_c})+B_2v_{2\beta}(v_1\cdot\varepsilon^\ast_{\chi_{c1}})(v_2\cdot
\varepsilon^\ast_{h_c})+B_3\varepsilon^{\ast}_{\chi_{c1}\beta}(v_2\cdot\varepsilon^\ast_{h_c})+
\end{equation}
\begin{displaymath}
+B_4\varepsilon^{\ast}_{h_c\beta}(v_1\cdot\varepsilon^\ast_{\chi_{c1}})+B_5v_{1\beta}
(\varepsilon^\ast_{\chi_{c1}}\cdot\varepsilon^\ast_{h_c})+
B_6v_{2\beta}(\varepsilon^\ast_{h_c}\cdot\varepsilon^\ast_{\chi_{c1}}),
\end{displaymath}
\begin{equation}
S_{2,\beta}(h_c+\chi_{c2})=\varepsilon^\ast_{\alpha\gamma}\Bigl[C_1\varepsilon_{\sigma\rho\beta\gamma}
v_1^\alpha v_1^\sigma v_2^\rho(v_2\cdot\varepsilon^\ast_{h_c})+
C_2\varepsilon_{\sigma\rho\beta\gamma}v_1^\alpha v_2^\sigma \varepsilon^{\ast\rho}_{h_c}+
C_3\varepsilon_{\sigma\rho\beta\gamma}\varepsilon^{\ast\alpha}_{h_c}v_1^\sigma v_2^\rho+
\end{equation}
\begin{displaymath}
+C_4g_{\alpha\beta}\varepsilon_{\sigma\rho\omega\gamma}v_1^\sigma v_2^\rho
\varepsilon^{\ast\omega}_{h_c}
+C_5\varepsilon_{\sigma\rho\lambda\beta}v_1^\alpha v_1^\gamma v_1^\sigma v_2^\rho
\varepsilon_{h_c}^{\ast\lambda}+C_6\varepsilon_{\sigma\rho\lambda\gamma}v_1^\alpha v_1^\sigma
v_{1\beta} v_2^\rho \varepsilon^{\ast\lambda}_{h_c}+,
\end{displaymath}
\begin{displaymath}
+C_7\varepsilon_{\sigma\rho\lambda\gamma}v_1^\alpha v_1^\sigma v_2^\rho v_{2\beta}
\varepsilon^{\ast\lambda}_{h_c}+C_8\varepsilon_{\sigma\lambda\beta\gamma}v_1^\alpha v_1^\sigma
\varepsilon^{\ast\lambda}_{h_c}\Bigr],
\end{displaymath}
where the coefficients $A_i$, $B_i$, $C_i$  can be presented
as sums of terms containing the factors $u=M_{\chi_{cJ}}/(M_{h_c}+M_{\chi_{cJ}})$,
$\kappa=m/(M_{h_c}+M_{\chi_{cJ}})$ and
$C_{ij}=c^i(p)c^j(q)=[(m-\epsilon(p))/(m+\epsilon(p))]^i
[(m-\epsilon(q))/(m+\epsilon(q))]^j$, preserving terms with
$i+j\leq 2$, and $r^2=(M_{h_c}+M_{\chi_{cJ}})^2/s$.
Exact analytical expressions for these coefficients
are sufficiently lengthy (compare with the results written in
Appendix A of our previous paper \cite{EFGM2009}), so we present them in
Appendix A of this work only in approximate numerical form using
the observed meson masses and the $c$-quark mass $m=1.55$ GeV.

Introducing the scattering angle $\theta$ between the electron
momentum ${\bf p}_e$ and the momentum ${\bf P}$ of the $h_c$
meson, we can calculate the differential cross section
$d\sigma/d\cos\theta$ and then the total cross section $\sigma$ as a
function of $r^2$. We find it useful to present the charmonium
production cross sections in the following form ($k=0,1,2$
corresponds to $\chi_{c0}$, $\chi_{c1}$ and $\chi_{c2}$):

\begin{equation}
\sigma(h_c+\chi_{cJ})=\frac{2\alpha^2\alpha_s^2(4m^2){\cal
Q}_c^2\pi
r^6\sqrt{1-r^2}\sqrt{1-r^2(2u-1)^2}}{9\kappa^4u^{11}(1-u)^{11}}\frac{|\tilde
R'_{h_c}(0)|^2|\tilde R'_{\chi_{cJ}}(0)|^2}{s(M_{\chi_{cJ}}+M_
{h_c})^{10}}\sum_{i=0}^7g^{(k)}F_i^{(k)}(r^2)\omega_i,
\end{equation}
where the functions $F_i^{(k)}$ ($k=0,1,2$) are written explicitly
in Appendix B. The factors $g^{(0)}=u^4(1-u)^2$, $g^{(1)}=\kappa^2/r^2$,
$g^{(2)}=1/16$ are introduced for the convenience,
\begin{equation}
\tilde R'_{\cal P}(0)=\frac{1}{3}\sqrt{\frac{2}{\pi}}\int_0^\infty
q^3R_{\cal P}(q)\frac{(\epsilon(q)+m)}{2\epsilon(q)}dq.
\end{equation}
The parameters $\omega_i$ can be expressed in terms of momentum
integrals $J_n$ for the states $h_c$ and $\chi_{cJ}$ as follows:
\begin{equation}
J_n=\int_0^\infty q^3R_{\cal
P}(q)\frac{(\epsilon(q)+m)}{2\epsilon(q)}\left(\frac{m-\epsilon(q)}
{m+\epsilon(q)}\right)^ndq,
\end{equation}
\begin{equation}
\omega_0=1,~~ \omega_1=\frac{J_1(h_c)}{J_0(h_c)},~~
\omega_2=\frac{J_2(h_c)}{J_0(h_c)},~~ \omega_3=\omega_1^2,
\end{equation}
\begin{displaymath}
\omega_4=\frac{J_1(\chi_{cJ})}{J_0(\chi_{cJ})},~~ \omega_5=\frac{J_2(\chi_{cJ})}{J_0(\chi_{cJ})},~~
\omega_6=\omega_4^2, ~~\omega_7=\omega_1\omega_4.
\end{displaymath}

On the one side, in the potential quark model the relativistic
corrections, connected with the relative motion of heavy $c$-quarks,
enter the production amplitude (2) and the cross section (19)
through the different relativistic factors. They are determined in
the final expression (19) by the specific parameters $\omega_i$. The
momentum integrals which determine the parameters $\omega_i$ are
convergent and we calculate them numerically, using the wave
functions obtained by the numerical solution of the Schr\"odinger
equation. The exact form of the wave functions $\Psi_0^{h_c}({\bf
p})$ and $\Psi_0^{\chi_{cJ}}({\bf q})$ is important for improving the
accuracy of the calculation of the relativistic effects. It is
sufficient to note that the double charmonium production cross
section $\sigma(s)$ in the nonrelativistic approximation contains
the factor $|R'_{h_c}(0)|^2 |R'_{\chi_{cJ}}(0)|^2$. Small changes of
the numerical values of the bound state wave functions at the origin
lead to substantial changes of the final results. In the framework of
NRQCD this problem is closely related to the
determination of the color-singlet matrix elements for the
charmonium \cite{BBL}. Thus, on the other side, there are
relativistic corrections to the bound state wave functions
$\Psi_0^{h_c}({\bf p})$, $\Psi_0^{\chi_{cJ}}({\bf q})$. In order to
take them into account, we suppose that the dynamics of a $c\bar
c$-pair is determined by the QCD generalization of the standard
Breit Hamiltonian in the center-of-mass reference frame \cite{repko1,pot1,pot3}:
\begin{equation}
H=H_0+\Delta U_1+\Delta U_2,~~~H_0=2\sqrt{{\bf
p}^2+m^2}-2m-\frac{C_F\tilde\alpha_s}{r}+Ar+B,
\end{equation}
\begin{equation}
\Delta U_1(r)=-\frac{C_F\alpha_s^2}{4\pi r}\left[2\beta_0\ln(\mu
r)+a_1+2\gamma_E\beta_0
\right],~~a_1=\frac{31}{3}-\frac{10}{9}n_f,~~\beta_0=11-\frac{2}{3}n_f,
\end{equation}
\begin{equation}
\Delta U_2(r)=-\frac{C_F\alpha_s}{2m^2r}\left[{\bf p}^2+\frac{{\bf
r}({\bf r}{\bf p}){\bf p}}{r^2}\right]+\frac{\pi
C_F\alpha_s}{m^2}\delta({\bf r})+\frac{3C_F\alpha_s}{2m^2r^3}({\bf
S}{\bf L})-
\end{equation}
\begin{displaymath}
-\frac{C_F\alpha_s}{2m^2}\left[\frac{{\bf S}^2}{r^3}-3\frac{({\bf
S}{\bf r})^2}{r^5}-\frac{4\pi}{3}(2{\bf S}^2-3)\delta({\bf
r})\right]-\frac{C_AC_F\alpha_s^2}{2mr^2},
\end{displaymath}
where ${\bf L}=[{\bf r}\times{\bf p}]$, ${\bf S}={\bf S}_1+{\bf S}_2$,
$n_f$ is the number of flavors, $C_A=3$ and $C_F=4/3$ are the
color factors of the SU(3) color group, $\gamma_E\approx 0.577216$ is
the Euler constant. To describe the hyperfine
splittings in ${\cal P}$-wave charmonium we add to the standard Breit potential
the scalar-exchange and vector-exchange confining potentials obtained
in \cite{repko4,gupta1987,repko2}:
\begin{equation}
\Delta V^{hfs}_{conf}(r)=
f_V\left[\frac{A}{2m^2r}\left(1+\frac{8}{3}{\bf S}_1 {\bf S}_2\right)+
\frac{3A}{2m^2r}{\bf L} {\bf S}+\frac{A}{3m^2r}\left(\frac{3}{r^2}({\bf S}_1 {\bf r}) ({\bf S}_2 {\bf r})-
{\bf S}_1 {\bf S}_2\right)\right]-
\end{equation}
\begin{displaymath}
-(1-f_V)\frac{A}{2m^2r}{\bf L} {\bf S},
\end{displaymath}
where we take the parameter $f_V=0.7$ for optimal agreement with
the experiment. For the dependence of the
QCD coupling constant $\tilde\alpha_s(\mu^2)$ on the renormalization point
$\mu^2$ in the pure Coulomb term in (23) we use the three-loop result \cite{kniehl1997}
\begin{equation}
\tilde\alpha_s(\mu^2)=\frac{4\pi}{\beta_0L}-\frac{16\pi b_1\ln L}{(\beta_0 L)^2}+\frac{64\pi}{(\beta_0L)^3}
\left[b_1^2(\ln^2 L-\ln L-1)+b_2\right], \quad L=\ln(\mu^2/\Lambda^2),
\end{equation}
whereas in other terms of the Hamiltonians (24) and (25) we take
the leading order approximation. The typical momentum transfer scale in a
quarkonium is of order of
the quark mass, so we set the renormalization scale $\mu=m$ and
$\Lambda=0.168$ GeV, which gives $\alpha_s=0.314$ for the charmonium
states. The coefficients $b_i$ are written explicitly in \cite{kniehl1997}.
The parameters of the linear potential $A=0.18$ GeV$^2$ and
$B=-0.16$ GeV have the usual values of quark models. Starting with the
Hamiltonian (23) we construct the effective potential model based on
the Schr\"odinger equation and find its numerical
solutions in the case of ${\cal P}$-wave charmonium \cite{LS}. The details of the used
model are presented in Appendix C. Then we calculate the matrix
elements entering in the expressions for the parameters $\omega_i$ (22)
and obtain the value of the production cross sections at
$\sqrt{s}=10.6$ GeV. Basic parameters which determine our numerical
results are collected in Table I. The comparison of the obtained
results with the previous calculations \cite{BL1,Chao,BLL1,Chao2008}
and experimental data \cite{Belle,BaBar} is presented in Table II.

\begin{table}
\caption{Numerical values of the relativistic parameters (20), (21),
(22) in the double charmonium production cross section (19).}
\bigskip
\begin{ruledtabular}
\begin{tabular}{|c|c|c|c|c|c|c|}
~~~Meson  $(c\bar c)$~~~&~~~ $n^{2S+1}L_J$ ~~~&~~ $J^{PC}$ ~~&~~ $M^{exp}$, GeV ~~&~~$\tilde R'_{\cal P}(0)$, GeV$^{5/2}$ ~~&
~~$\omega_1$ or $\omega_4$~~&~~ $\omega_2$ or $\omega_5$~~\\
\hline
$\chi_{c0}$&$1^3P_0$&$0^{++}$  & 3.415 & 0.33 & -0.28  & 0.13
\\  \hline
$\chi_{c1}$&$1^3P_1$&$1^{++}$   & 3.511 & 0.20 & -0.18  & 0.07
\\  \hline
$\chi_{c2}$&$1^3P_2$&$2^{++}$   & 3.556 & 0.13 & -0.08  & 0.01
\\  \hline
$h_c$&$1^1P_1$&$1^{+-}$   & 3.525 & 0.17 & -0.14  & 0.04
\\  \hline
\end{tabular}
\end{ruledtabular}
\end{table}

\section{Numerical results and discussion}

In this paper we have investigated the role of relativistic effects
in the production processes of ${\cal P}$-wave mesons $(c\bar c)$ in
the quark model. At the calculation of the production amplitude
(2) we keep relativistic corrections of two types. The first type is
determined by several functions depending on the relative quark
momenta  ${\bf p}$ and ${\bf q}$ arising from the gluon propagator,
the quark propagator and the relativistic meson wave functions. The
second type of corrections originates from the perturbative and nonperturbative
treatment of the quark-antiquark interaction operator which leads to
the different wave functions $\Psi_0^{h_c}({\bf p})$ and
$\Psi_0^{\chi_{cJ}}({\bf q})$ for the ${\cal P}$-wave charmonium
states. In addition, we systematically accounted for
the bound state corrections working with the observed masses of
${\cal P}$-wave mesons ($\chi_{cJ}$,
$h_c$). The calculated masses of ${\cal P}$-wave charmonium states
agree well with experimental values \cite{PDG} (see Table III).
Note that the basic parameters of the model are kept fixed from the
previous calculations of the meson mass spectra and decay widths
\cite{rqm5,rqm1,brambilla-2011,QWG}. The strong coupling constant entering the
production amplitude (2) is taken to be $\alpha_s=0.24$ in
accordance with the leading order QCD relation at $\mu=2m$.

Numerical results and their comparison with the previous
calculation in NRQCD are presented in Table II.
We have included in it also new numerical results (several numerical
mistakes contained in \cite{EM2010} were corrected) obtained on the
basis of quark model (23)-(26) for the production cross sections
of a pair of $S$- and ${\cal P}$-wave charmonium states.
The exclusive double charmonium production cross section presented
in the form (19) is convenient for a comparison with the results of
NRQCD. Indeed, in the nonrelativistic limit, when $u=1/2$,
$\kappa=1/4$, $\omega_i=0$ ($i\geq 1$), $r^2=16m^2/s$, the cross
section (19) coincides with the calculation in \cite{BL1}. In this
limit the functions $F_0^{(k)}(r^2)$ transform into corresponding
functions $F_k$ from \cite{BL1}. When we take into account bound
state corrections working with observed meson masses, we get
$u=M_{\chi_{cJ}}/(M_{h_c}+M_{\chi_{cJ}}) \not = 1/2$,
$\kappa=m/(M_{h_c}+M_{\chi_{cJ}})\not =1/4$. This leads to the
modification of the general factor in (19) and the form of the
functions $F_0^{(k)}$ in comparison with the nonrelativistic theory
(see \cite{BL1}). It follows from the numerical values of the
parameters $\omega_i$, presented in Table I, that the relativistic
corrections could amount to $10\div 30 \%$ in the production amplitude.
In fact their influence on the value of the production cross sections
become considerably larger in the case of reactions $e^++e^-\to
h_c+\chi_{c1,c2}$. Only due to relativistic contributions to the
production amplitude the cross section $\sigma(e^++e^-\to h_c+\chi_{c1})$
increases in two times and $\sigma(e^++e^-\to h_c+\chi_{c2})$ in four times
in comparison with the nonrelativistic calculation. Opposite influence
on the value of the cross sections is determined by relativistic corrections
to the bound state wave functions in the rest frame.
Indeed, relativistic effects change considerably the values of the
nonrelativistic parameters $R'_{\cal P}(0)$, which transform into
$\tilde R'_{\cal P}(0)$ (20). Different values of the mass
of $c$-quark and nonperturbative parameters $R'_{\cal P}(0)$ make
difficult the direct comparison of our numerical results with predictions
of NRQCD. Note that nonrelativistic results obtained in our quark model
are the following: $\sigma(\chi_{c0}+h_c)=0.101$ fb, $\sigma(\chi_{c1}+h_c)=0.417$ fb,
$\sigma(\chi_{c2}+h_c)=0.026$ fb (compare with predictions of NRQCD in
fourth column of Table II).
Nevertheless, we can state that in all considered
reactions $e^++e^-\to h_c+\chi_{cJ}$ the account of all
relativistic effects leads to the decrease of the nonrelativistic cross section obtained
in our model. It is necessary to point out once again that the essential
effect on the value of the production cross sections
$h_c+\chi_{cJ}$ belongs to the parameters $\tilde R'_{\cal P}(0)$ (20), $\alpha_s$, $m$.
Small changes in their values can lead to significant changes in the
production cross sections. In our model the nonrelativistic value
$R'_{\cal P}(0)=0.24$ GeV$^{5/2}$. Accounting for the potentials (23)-(26) which give
the good mass splitting for ${\cal P}$-wave charmonium states, we observe
simultaneously the decreasing and splitting in the parameter $\tilde R'_{\cal P}(0)$
(see Table I). As a result the nonrelativistic cross sections
$\sigma(e^++e^-\to h_c+\chi_{c1})$ and $\sigma(e^++e^-\to h_c+\chi_{c2})$
decrease in three and six times correspondingly and $\sigma(e^++e^-\to h_c+\chi_{c0})$ reduces
approximately on $25\%$.

We presented a systematic treatment of relativistic effects in the
${\cal P}$-wave double charmonium production in $e^+e^-$ annihilation.
We separated two different types of relativistic
contributions to the production amplitudes. The first type includes
the relativistic $v/c$ corrections to the wave functions and their
relativistic transformations. The second type includes the
relativistic $p/\sqrt{s}$ corrections appearing from the expansion of
the quark and gluon propagators. The latter corrections were taken
into account up to the second order. It is important to note that
the expansion parameter $p/\sqrt{s}$ is very small. In our analysis
of the production amplitudes we correctly take into account
relativistic contributions of order $O(v^2/c^2)$ for
the ${\cal P}$-wave mesons. Therefore the first basic
theoretical uncertainty of our calculation is connected with the
omitted terms of order $O({\bf p}^4/m^4)$. Since the calculation
of the masses of ${\cal P}$-wave charmonium states is sufficiently accurate in our
model (the error is less then 1 $\%$), we suppose
that the uncertainty in the cross section calculation due to the
omitted relativistic corrections of order $O({\bf p}^4/m^4)$ in the
quark interaction operator (the Breit Hamiltonian) is also very small.
Taking into account that
the average value of the heavy quark velocity squared in the
charmonium is $\langle v^2\rangle=0.3$, we expect that relativistic corrections
of order $O({\bf p}^4/m^4)$ should not exceed $30\%$ of the obtained
relativistic contribution. Strictly speaking in the quasipotential approach
we can not find precisely the bound
state wave functions in the region of the relativistic momenta $p\ge m$ which gives
near $30\%$ of the total value $\sigma$ (19). Using indirect arguments related with the
mass spectrum calculation we estimate in $10\%$ the uncertainty in the wave function
determination. Larger value of the error will lead to the essential
discrepancy between the experiment and theory in the calculation of the charmonium mass
spectrum. Then the corresponding error in the cross section (19) is not exceeding $15\%$.
The significant improvement in the calculation of the relativistic corrections
to the double charmonium cross section $\sigma(e^+e^-\to J/\Psi+\eta_c)$ was obtained
in \cite{bodwin2007} in the nonrelativistic QCD factorization formalism. The essential
refinement was connected with many factors including the resummation of a class of
relativistic corrections and the contribution that arises from the interference
between the relativistic corrections  and the corrections of the next to leading order
in $\alpha_s$. In our work the appearance of divergent integrals over ${\bf p}$ and
${\bf q}$ for the corrections of order $O({\bf p}^4/m^4)$ and $O({\bf q}^4/m^4)$ is
the consequence of expansions (10)-(12) used by us in order to perform analytically
the angular integration in (2). The omitted corrections can be included and the obtained results
can be improved if we calculate all integrals over the relative momenta ${\bf p}$ and
${\bf q}$ in (2) without any expansions.
Another important part of the total theoretical error is related with
radiative corrections of order $\alpha_s$ which were omitted in our analysis.
Our approach to the calculation of the amplitude of the double charmonium production
can be extended beyond the leading order in the strong coupling constant. Then the vertex
functions in (2) will have more complicate structure including the integration over the
loop momenta. Our calculation of the cross
sections accounts for effectively only some part of one loop corrections by means of the
Breit Hamiltonian. So, we assume that the radiative corrections of order $O(\alpha_s)$
can cause $20\%$ modification of the production cross sections.
We have neglected the terms in
the cross section (19) containing the product of $J_n$
with summary index $>2$ because their contribution has been found
negligibly small. There are no another comparable uncertainties
related to the other parameters of the model,
since their values were fixed from our previous consideration of
meson and baryon properties \cite{rqm1,rqm5}. Our total theoretical errors are
written explicitly in Table II. To obtain this estimate we add the above
mentioned uncertainties in quadrature.

\begin{table}
\caption{Comparison of the obtained results with previous
theoretical predictions and experimental data.}
\bigskip
\begin{ruledtabular}
\begin{tabular}{|c|c|c|c|c|c|c|c|}
State  & $\sigma_{BABAR}\times$ & $\sigma_{Belle}\times $
&$\sigma_{NRQCD}$& $\sigma$ (fb) &$\sigma$ (fb) & $\sigma$
(fb) & Our result \\
$H_1H_2$   &$ Br_{H_2\to charged\ge 2}$ & $Br_{H_2\to charged\ge 2}$
&(fb) \cite{BL1}& \cite{Chao}  &\cite{BLL1}& \cite{Chao2008}  &  (fb)  \\
 &(fb) \cite{BaBar} & (fb) \cite{Belle}    &  &  &    &    &
\\   \hline
$J/\Psi+\chi_{c0}$ & $10.3\pm 2.5^{+1.4}_{-1.8}$ & $6.4\pm
1.7\pm 1.0$ &  $2.40\pm 1.02$& 6.7& 14.4  & 17.9(6.35)  &  $14.47\pm 5.64$ \\
\hline  $J/\Psi+\chi_{c1}$ &  & & $0.38\pm 0.12$& 1.1 &  & &
$1.78\pm 0.69$ \\  \hline $J/\Psi+\chi_{c2}$ &  & &  $0.69\pm 0.13$ &1.6 & & & $0.44\pm 0.17$  \\
\hline $\eta_c+h_c$ & &  & $0.308\pm 0.017$&  & &  &  $0.25\pm 0.10$ \\  \hline
$h_c+\chi_{c0}$ & &  & $0.053\pm 0.019$&  & &  &  $0.075\pm 0.029$ \\  \hline
$h_c+\chi_{c1}$ & &  &$0.258\pm 0.064$ &  & &  &  $0.132\pm 0.051$ \\  \hline
$h_c+\chi_{c2}$ & &  & $0.017\pm 0.002$&  & &  &  $0.004\pm 0.002$ \\  \hline
\end{tabular}
\end{ruledtabular}
\end{table}

\acknowledgments

The authors are grateful to D. Ebert, R.N. Faustov and V.O. Galkin
for useful discussions. The work is performed under the
financial support of the Federal Program "Scientific and pedagogical
personnel of innovative Russia"(grant No. NK-20P/1).

\appendix

\section{The coefficients $A_i$, $B_i$ and $C_i$ entering in
the production amplitudes (16)-(18)}

These coefficients are the sums of the terms containing the
parameters $u=M_{\chi_{cJ}}/(M_{h_c}+M_{\chi_{cJ}})$ and
$\kappa=m/(M_{h_c}+M_{\chi_{cJ}})$. We present $A_i$, $B_i$ and $C_i$
in numerical form using the observed meson masses and the mass
of $c$-quark $m=1.55$ GeV.\\[1mm]

\vspace{1mm}
{\underline {$e^++e^-\to h_c+\chi_{c0}$}}

\begin{equation}
A_0=47.20-6.63r^2+C_{01}(-4.66+37.70r^2-8.05r^4+0.01r^6)+
\end{equation}
\begin{displaymath}
+C_{10}(-40.36+55.54r^2-6.32r^4+0.01r^6)+6.63r^2C_{02}+C_{20}(3.24+5.00r^2)+
\end{displaymath}
\begin{displaymath}
+C_{11}(3.98-48.83r^2+53.37r^4-7.63r^6+0.02r^8).
\end{displaymath}

\vspace{3mm}
{\underline {$e^++e^-\to h_c+\chi_{c1}$}}

\begin{equation}
B_1=-4r^2+2.73r^4+C_{10}(3.42r^2-14.22r^4+3.68r^6)+C_{20}(4.44r^2-2.73r^4)+
\end{equation}
\begin{displaymath}
+C_{01}(-2.75r^2-11.24r^4+3.89r^6)+C_{11}(2.35r^2+25.47r^4-26.77r^6+5.06r^8)-2.73r^4C_{02},
\end{displaymath}
\begin{equation}
B_2=-7.50r^2+2.73r^4+C_{10}(24.13 r^2-19.67r^4+3.68r^6)+C_{20}(7.50r^2-2.73r^4)+
\end{equation}
\begin{displaymath}
+C_{01}(16.45r^2-17.04r^4+3.89r^6)+C_{02}(4.51r^2-2.73r^4)+C_{11}(-44.63r^2+65.23r^4-34.83r^6+5.06r^8),
\end{displaymath}
\begin{equation}
B_3=18.99-9.19r^2+C_{10}(-51.66+58.38r^2-13.28r^4)+C_{20}(-18.99+8.75r^2)+
\end{equation}
\begin{displaymath}
+C_{01}(-28.91+49.12r^2-13.49r^4)+C_{02}(-9.03+8.21r^2)+
\end{displaymath}
\begin{displaymath}
+C_{11}(85.84-164.44r^2+109.27r^4-17.99r^6),
\end{displaymath}
\begin{equation}
B_4=4-1.82r^2+C_{10}(-3.42+9.48r^2-1.84r^4)+C_{20}(-2+1.82r^2)+
\end{equation}
\begin{displaymath}
+C_{01}(1.51+7.53r^2-1.95r^4)+1.82r^2C_{02}+C_{11}(-1.29-17.07r^2+13.40r^4-2.03r^6),
\end{displaymath}
\begin{equation}
B_5=C_{01}(2.50-0.12r^2+0.02r^4)+C_{20}(-2.88+0.44r^2)+C_{11}(-2.13+0.28r^2-0.08r^4-0.05r^6),
\end{equation}
\begin{equation}
B_6=C_{01}(-2.48+0.06r^2+0.02r^4)+0.44r^2C_{20}+C_{11}(2.12-0.22r^2+0.14r^4-0.05r^6).
\end{equation}

\vspace{3mm}
{\underline {$e^++e^-\to h_c+\chi_{c2}$}}

\begin{equation}
C_1=-2.36r^4+C_{10}(8.74r^4-3.28r^6)+2.36r^4C_{20}+C_{01}(0.35r^2+10.22r^4-3.22r^6)+
\end{equation}
\begin{displaymath}
+C_{11}(-0.42r^2-31.78r^4+24.24r^6-4.33r^8)+2.36r^4C_{02},
\end{displaymath}
\begin{equation}
C_2=-1.21+C_{10}(1.67-0.78r^2)+0.42C_{20}+0.82C_{02}+
\end{equation}
\begin{displaymath}
+C_{01}(1.16+0.46r^2+0.41r^4)+C_{11}(-1.16-1.50r^2-0.71r^4+0.66r^6),
\end{displaymath}
\begin{equation}
C_3=-1.21+1.58r^2+C_{10}(1.67-6.60r^2+1.64r^4)+C_{20}(0.42-1.58r^2)+C_{02}(0.82-1.58r^2)+
\end{equation}
\begin{displaymath}
+C_{01}(0.46-6.65r^2+1.61r^4)+C_{11}(-0.33+19.72r^2-11.80r^4+1.73r^6),
\end{displaymath}
\begin{equation}
C_4=1.21+C_{10}(-1.67+0.78r^2)-0.42C_{20}+C_{11}(0.33+1.70r^2-0.60r^4)+
\end{equation}
\begin{displaymath}
+C_{01}(-0.46-0.38r^2)-0.82C_{02},
\end{displaymath}
\begin{equation}
C_5=C_{11}(-0.58r^4+0.89r^6)+0.81r^4C_{01},
\end{equation}
\begin{equation}
C_6=-0.12r^4C_{01}+C_{11}(-0.01r^4-0.18r^6),
\end{equation}
\begin{equation}
C_7=C_{01}(-0.35r^2+0.33r^4)+C_{11}(0.42r^2-0.35r^4+0.55r^6),
\end{equation}
\begin{equation}
C_8=C_{01}(0.47r^2+0.40r^4)+C_{11}(-0.51r^2-0.57r^4+0.73r^6)-0.38r^2C_{20}.
\end{equation}

\section{The functions $F_i^{(k)}(r^2)$ ($k=0,1,2$) entering in the
production cross section (19)}

{\underline {$e^++e^-\to h_c+\chi_{c0}$}}

\begin{equation}
F_0^{(0)}=2.25-2.88r^2+0.68r^4-0.04r^6,
\end{equation}
\begin{equation}
F_1^{(0)}=-3.84+9.67r^2-7.17r^4+1.43r^6-0.09r^8,
\end{equation}
\begin{equation}
F_2^{(0)}=0.31+0.12r^2-0.50r^4+0.07r^6,
\end{equation}
\begin{equation}
F_3^{(0)}=1.64-6.16r^2+8.14r^4-4.33r^6+0.75r^8,
\end{equation}
\begin{equation}
F_4^{(0)}=-0.44+4.09r^2-4.92r^4+1.38r^6-0.11r^8,
\end{equation}
\begin{equation}
F_5^{(0)}=0.63r^2-0.72r^4+0.09r^6,
\end{equation}
\begin{equation}
F_6^{(0)}=0.02-0.38r^2+1.86r^4-2.12r^6+0.68r^8,
\end{equation}
\begin{equation}
F_7^{(0)}=0.76-9.05r^2+18.96r^4-13.49r^6+3.03r^8.
\end{equation}

\vspace{1mm}
{\underline {$e^++e^-\to h_c+\chi_{c1}$}}

\begin{equation}
F_0^{(1)}=0.19+2.00r^2-4.32r^4+2.64r^6-0.51r^8,
\end{equation}
\begin{equation}
F_1^{(1)}=-0.32-11.48r^2+30.85r^4-28.58r^6+11.00r^8-1.47r^{10},
\end{equation}
\begin{equation}
F_2^{(1)}=-0.37-3.96r^2+8.45r^4-5.10r^6+0.98r^8,
\end{equation}
\begin{equation}
F_3^{(1)}=0.14+15.71r^2-51.51r^4+64.29r^6-37.91r^8+10.34r^{10},
\end{equation}
\begin{equation}
F_4^{(1)}=0.14-6.78r^2+21.01r^4-22.91r^6+10.03r^8-1.49r^{10},
\end{equation}
\begin{equation}
F_5^{(1)}=-2.13r^2+5.02r^4-3.81r^6+0.92r^8,
\end{equation}
\begin{equation}
F_6^{(1)}=0.03+4.78r^2-21.19r^4+35.38r^6-26.94r^8+9.03r^{10},
\end{equation}
\begin{equation}
F_7^{(1)}=-0.24+38.00r^2-133.92r^4+186.34r^6-123.58r^8+37.52r^{10}.
\end{equation}

\vspace{1mm}
{\underline {$e^++e^-\to h_c+\chi_{c2}$}}

\begin{equation}
F_0^{(2)}=0.23-0.81r^2+0.98r^4-0.40r^6,
\end{equation}
\begin{equation}
F_1^{(2)}=-1.72+6.75r^2-9.73r^4+5.83r^6-1.13r^8,
\end{equation}
\begin{equation}
F_2^{(2)}=-0.47+1.61r^2-1.91r^4+0.76r^6,
\end{equation}
\begin{equation}
F_3^{(2)}=3.18-13.83r^2+23.15r^4-18.01r^6+6.32r^8,
\end{equation}
\begin{equation}
F_4^{(2)}=-1.92+7.56r^2-10.88r^4+6.34r^6-1.11r^8,
\end{equation}
\begin{equation}
F_5^{(2)}=-0.47+1.63r^2-1.96r^4+0.80r^6,
\end{equation}
\begin{equation}
F_6^{(2)}=4.14-17.68r^2+29.23r^4-21.88r^6+6.98r^8,
\end{equation}
\begin{equation}
F_7^{(2)}=13.29-58.23r^2+97.89r^4-75.59r^6+25.74r^8.
\end{equation}

\section{Effective relativistic Hamiltonian}

For the calculation of the relativistic corrections in the bound
state wave functions $\Psi_0^{\cal P}$ we
consider the Breit potential (23). It contains a number of terms
which should be transformed in order to use the program of numerical
solution of the Schr\"odinger equation \cite{LS}. The
rationalization of the kinetic energy operator can be done in the
following form \cite{Lucha}:
\begin{equation}
T=2\sqrt{{\bf p}^2+m^2}=2\frac{{\bf p}^2+m^2}{\sqrt{{\bf
p}^2+m^2}}\approx \frac{{\bf p}^2}{\tilde m}+\frac{2m^2}{\tilde E},
\end{equation}
where $\tilde m$ is the effective mass of heavy quarks,
\begin{equation}
\tilde m=\frac{\tilde E}{2}=\frac{\sqrt{{\bf p}^2_{eff}+m^2}}{2}.
\end{equation}
${\bf p}^2_{eff}$ should be considered as a new parameter which
effectively accounts for relativistic corrections in (C1). We take numerical
value of ${\bf p}^2_{eff}=0.54$ GeV$^2$ for ${\cal P}$-wave charmonium states (see Table
III). The second term in the
Breit potential (23), which also has to be transformed, takes the
form:
\begin{equation}
\Delta\tilde U=-\frac{2\alpha_s}{3m^2r}\left[{\bf
p}^2-\frac{d^2}{dr^2}\right].
\end{equation}
It has the similar structure as the operator of effective kinetic energy from
the Hamiltonian $H_0$. So, we change slightly the code of the Mathematica
programm in \cite{LS} in order to include the correction
$\Delta\tilde U$ directly in the initial Hamiltonian.

\begin{table}
\caption{The parameters of the effective relativistic Hamiltonian and masses of
${\cal P}$-wave charmonium states.}
\bigskip
\begin{ruledtabular}
\begin{tabular}{|c|c|c|c|c|c|}
~~~Meson $(c\bar c)$~~~&~~~$n^{2S+1}L_J$ ~~~&~~~${\bf p}^2_{eff}$, GeV$^2$ ~~~&
~~~$\tilde m$, GeV~~~&~~~ $M^{th}$, GeV~~~&~~~ $M^{exp}$, GeV  \cite{PDG}~~~ \\  \hline
$\chi_{c0}$&$1^3P_0$ &0.54 & 0.857&  3.418& 3.415\\
\hline $\chi_{c1}$&$1^3P_1$  & 0.54 & 0.857 &
3.493&
3.511\\  \hline $\chi_{c2}$&$1^3P_2$  & 0.54& 0.857& 3.557& 3.556 \\
\hline $h_c$&$1^1P_1$  &0.54& 0.857 & 3.499& 3.525\\
\hline
\end{tabular}
\end{ruledtabular}
\end{table}

At last, there is a need to transform the spin-spin
and spin-orbit interactions in (25) which have the $1/r^3$ behavior at
small $r$. For the purpose of solving the Schr\"odinger equation we consider the
regularization of such terms due to the account of
the relative motion of heavy quarks which was discussed many times in
\cite{repko1,gupta1987,repko2}. The nonsingular potentials in both cases have
the following structure at small $r$: $(1-f_i)/r^3$ with $f_1=(1+2mr)e^{-2mr}$
and $f_2=(1+2mr+4m^2r^2/3)e^{-2mr}$ for spin-orbit and spin-spin interactions
correspondingly. In Table III we present the results of the
calculation of the ${\cal P}$-wave charmonium mass spectrum and a comparison with
the existing experimental data. The obtained masses agree with the
experimental ones within an accuracy $1\%$. So we can
suppose that the obtained effective Hamiltonian allows to account
relativistic corrections in the bound state wave functions with
sufficiently good accuracy.

\end{document}